\documentclass{article}
\usepackage[hidelinks]{hyperref}

\usepackage[preprint]{spconf}
\usepackage{amsmath,graphicx}

\usepackage{url}
\usepackage{etoolbox}

\usepackage{booktabs}
\usepackage{siunitx}

\newcolumntype{H}{>{\setbox0=\hbox\bgroup}c<{\egroup}@{}}
\newrobustcmd\B{\DeclareFontSeriesDefault[rm]{bf}{b}\bfseries}

\usepackage[above,below]{placeins}

\usepackage{adjustbox}

\usepackage{microtype}

\usepackage{comment}

\usepackage{orcidlink}

\copyrightnotice{\copyright\ IEEE 2026}
\toappear{Submitted to {\it ICASSP 2026}}

\title{Quality Assessment of Noisy and Enhanced Speech with Limited Data: UWB-NTIS System for VoiceMOS 2024} 

\name{Marie Kunešová\orcidlink{0000-0002-7187-8481}, Aleš Pražák\orcidlink{0000-0001-9453-0034}, Jan Lehečka\orcidlink{0000-0002-3889-8069}}
\address{NTIS Research Centre, 
Faculty of Applied Sciences, University of West Bohemia in Pilsen, Czechia
}

\begin{document}
\ninept
\maketitle
\begin{abstract}
We present a system for non-intrusive prediction of speech quality in noisy and enhanced speech, developed for Track 3 of the VoiceMOS 2024 Challenge. The task required estimating the ITU-T P.835 metrics SIG, BAK, and OVRL without reference signals and with only 100 subjectively labeled utterances for training. Our approach uses wav2vec 2.0 with a two-stage transfer learning strategy: initial fine-tuning on automatically labeled noisy data, followed by adaptation to the challenge data. The system achieved the best performance on BAK prediction ($\text{LCC}=0.867$) and a very close second place in OVRL ($\text{LCC}=0.711$) in the official evaluation. Post-challenge experiments show that adding artificially degraded data to the first fine-tuning stage substantially improves SIG prediction, raising correlation with ground truth scores from 0.207 to 0.516. These results demonstrate that transfer learning with targeted data generation is effective for predicting P.835 scores under severe data constraints.
\end{abstract}
\section{Introduction}

There has long been an interest in the assessment of speech quality across different speech applications, as researchers strive to improve their speech processing systems. For instance, speech synthesis researchers aim for improving the intelligibility and naturalness of synthesized speech, while developers of speech enhancement systems wish to remove unwanted noise while avoiding audible signal distortion. Both groups regularly make use of subjective listening tests, but these can be both expensive and time-consuming, as well as requiring a great number of human participants~\cite{Wester2015}, and thus, automated alternatives are being sought.

However, the automatic assessment of speech quality is still far from a fully solved problem. This is what motivated the VoiceMOS challenges, a series of machine learning competitions aimed at this very task. VoiceMOS Challenge (VMC) 2024~\cite{huang2024voicemos2024} consisted of three tracks: speech quality assessment of 1) synthesized and voice converted speech, 2) synthesized, voice converted, and resynthesized singing voice, and 3) noisy and enhanced (de-noised) natural speech.

In the past, we have participated in all three editions of the VoiceMOS Challenge (2022, 2023, and 2024) as team UWB-NTIS-TTS~\cite{Kunesova2025VoiceMOS_journal}. 
However, until VMC 2024, our focus has been solely on the automatic prediction of the mean opinion score (MOS) \cite{ITU.P.800.1} of text-to-speech and voice conversion systems. 

In this paper, we instead focus on the last of the VMC 2024 tasks, presenting our system for the automatic prediction of audio quality in noisy and enhanced \emph{natural} speech. 
The system was never published during the challenge itself. This paper therefore serves as the first full description of the system and additionally presents extensions and evaluations that were not part of the challenge submission. 

\subsection{Track 3 of VoiceMOS Challenge 2024}
\label{sec:voicemos2024_track3}

In Track 3 of the VMC 2024 Challenge, participants were given samples of clean, noisy or enhanced (de-noised) \emph{natural speech} recordings and were tasked with predicting three quality measures established by the ITU-T P.835 recommendation for the evaluation of noise-suppression algorithms~\cite{ITU.P.835}: \emph{SIG}, measuring the speech quality and level of speech distortion; \emph{BAK}, measuring the level and intrusiveness of background noise; and \emph{OVRL}, measuring the overall quality of the speech sample.
These measures are very similar to MOS -- like it, they are obtained through listening tests, where human listeners grade utterances on a scale of 1--5, where 5 denotes the best quality.

In this track, participants also had to deal with a very restricting limitation: 
aside from the very small training and validation sets provided by the organizers (60 and 40 utterances, respectively, each with a duration of 4--5 seconds), no additional data \emph{with a subjective quality label} could be used for training. However, other types of data could still be used for data augmentation or pretraining.

The training and validation data for VMC 2024 Track 3 came from the CHiME 7-UDASE dataset~\cite{LEGLAIVE2025101685}, while the 280 files in the evaluation set were created by the VoiceMOS organizers from the VoiceBank-DEMAND~\cite{valentinibotinhao16_ssw} dataset. However, the origin of the data was only revealed after the conclusion of the challenge~\cite{huang2024voicemos2024}.

\subsection{Related work} 

There are many different approaches to automatic speech quality assessment. Objective metrics such as Perceptual Evaluation of Speech Quality (PESQ)~\cite{Rix2001} and Perceptual Objective Listening Quality Assessment (POLQA)~\cite{ITU.P.863} measure the distortion between a speech sample and a clean reference. However, such methods generally require access to the original signal.

Other automatic approaches focus on the prediction of subjective or objective metrics without a reference, typically using deep learning. In the speech synthesis and voice conversion domains, many works predict the mean opinion score (MOS)~\cite{ITU.P.800.1}, ordinarily obtained via listening tests, using methods such as long short-term memory networks (LSTM)~\cite{Patton2016}, convolutional neural networks (CNN)~\cite{Yoshimura2016}, or various self-supervised learning models~\cite{Saeki22c_UTMOS}. 

For the assessment of noisy and enhanced speech, some authors also focus on metrics derived from MOS, such as in~\cite{Avila2019}, using Mel Frequency features and DNN. Meanwhile, the system proposed in \cite{Dong2020} estimates several objective metrics, including PESQ, without a reference signal, while the authors of~\cite{deoliveira25_interspeech} introduce an unsupervised diffusion-based system which is not trained to predict any specific metric, but nevertheless correlates well with human ratings. Finally, the CNN-based NISQA~\cite{mittag21_interspeech} and transformer-based SQ-AST~\cite{wardah25_interspeech} systems evaluate both the overall speech quality and four additional aspects: noisiness, discontinuity, coloration, and loudness.

However, very few automatic speech quality assessment methods focus specifically on predicting ITU-T P.835 scores, which were the objective of the VMC 2024 Track 3. This may be in part due to a lack of publicly available training data, aside from the previously mentioned recently released CHiME 7 - UDASE dataset~\cite{LEGLAIVE2025101685}. One well-known exception is the CNN-based DNSMOS P.835 \cite{Reddy2022_DNSMOS} system, which was trained on an (to our best knowledge unreleased) dataset of 30,000 P.835-labeled audio samples. In comparison, the VMC 2024 set of 100 labeled utterances is very small indeed.

\section{System description}

The restrictions on available training data led us to choose an approach based on transfer learning. 
First, we take a pretrained speech representation model and fine-tune it for a different, related task, where data with ``non-subjective'' labels can be used. Secondly, we re-fine-tune it for the target task using the small amount of subjectively-labeled data provided in the challenge. To facilitate this, we treat each of the three predicted scores (SIG, BAK, and OVRL) independently, implementing a separate approach for each.

\begin{figure}[ht]
    \centering
    \includegraphics[width=\linewidth]{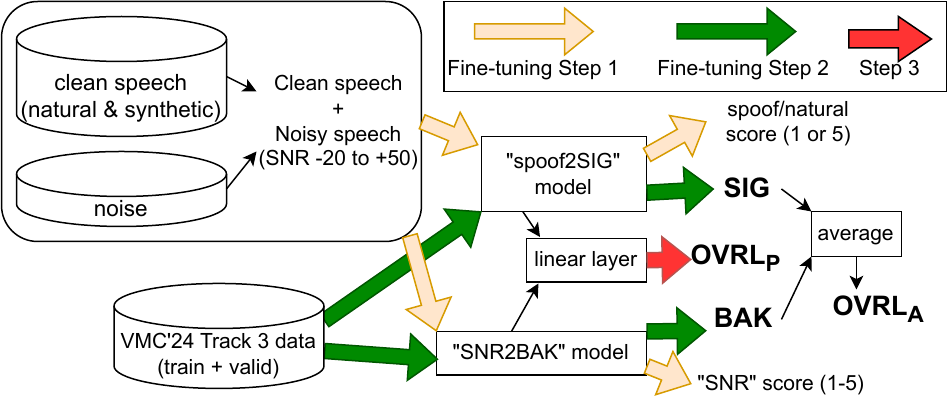}
    \caption{Diagram of our system, with both approaches to obtaining OVRL -- by averaging (OVRL$_A$) or prediction (OVRL$_P$).} 
    \label{fig:voicemos2024_track3}
\end{figure}

As seen in the system's diagram in Figure~\ref{fig:voicemos2024_track3}, we created two main prediction models, predicting SIG and BAK. For the OVRL scores, which rate the overall quality of each utterance and thus should combine both SIG and BAK, we considered two simpler approaches: either a) merely calculating the average of the predicted SIG and BAK values (OVRL$_A$), or b) combining the fine-tuned SIG and BAK models with a new head trained for OVRL prediction (OVRL$_P$). 

As the basis of our SIG and BAK prediction approach, we employed pretrained wav2vec 2.0 models~\cite{baevski2020wav2vec}. For our initial VMC 2024 submission, we considered two existing pretrained models: the most basic English model \emph{wav2vec2-base} \cite{baevski2020wav2vec} (``w2v2-base'') and the Czech model \emph{wav2vec2-base-cs-80k-ClTRUS} \cite{IS22_ASR_wav2vec2_Lehecka} (``ClTRUS''). The latter was chosen because it has the same number of parameters as w2v2-base, but is pretrained on much richer and noisier data, which seems more suitable for this task. However, as the VMC 2024 data consist of English utterances, there is a language mismatch -- we expect that this mismatch may negatively affect SIG prediction, but believe that it should not matter for BAK, which only focuses on the noise level of each utterance.

After the VoiceMOS Challenge, we also added a third wav2vec 2.0 model, ``w2v2-dgrd'', which we pretrained ourselves specifically for this task, using artificially degraded speech data. This process is described in Section~\ref{sec:w2v2-dgrd}.

The models for BAK and SIG prediction were fine-tuned in two stages, first on simulated noisy data and then on the VMC 2024 Track 3 training and validation data. In both cases, only the feature extractor was frozen during both stages of fine-tuning.
With the exception of the number of epochs, other training parameters were also identical for all models and both stages of fine-tuning: the models were trained for regression with MSE loss and mean-pooling, with a learning rate of $3 \times 10^{-5}$ and a learning rate warm-up ratio of 0.1. The models were trained for 20 epochs in the first stage and 300--700 epochs in the second stage, with an earlier checkpoint selected on an individual basis, based on validation loss. 
Because of the very short durations of the audio files (4 seconds for the training and validation data, 1--8 seconds for the evaluation data), the input length of the training data was set to 1 second, randomly sampled from each input file during training.

\subsection{Training data for model fine-tuning}
\label{sec:traindata}

For the first step of fine-tuning of both models, we generated noisy training data using a modified version of the Microsoft Scalable Noisy Speech Dataset (MS-SNSD) \cite{reddy2019scalable}, with additional noises and clean speech. For clean speech, we used:
\begin{itemize}
    \item The default data included in MS-SNSD, i.e., PTDB-TUG~\cite{pirker11_ptdbtug} 
    and the training set of the ``Edinburgh 56 speaker dataset'' 
    (also known as VoiceBank-DEMAND~\cite{valentinibotinhao16_ssw})\footnote{We have since learned that the VMC 2024 eval. set was \emph{also} created from the Edinburgh/VoiceBank-DEMAND dataset (but from the test set, which has different speakers). We were not aware of this at the time of the challenge, so the choice of training data from the same dataset 
    was purely coincidental.}.
    \item LibriSpeech~\cite{Panayotov_Librispeech}, using approximately 7000 shortest files from the ``train-clean-100'' set
    \item ASVSpoof2019~\cite{Wang_ASVspoof2019}, using all files from the training and evaluation sets of the \emph{logical access} (LA) scenario, including both natural and synthetic (``spoofed'') speech.
\end{itemize}

As noises, we used the default noises included in MS-SNSD, as well as additional noises from the ESC-50 dataset\footnote{downloaded from https://github.com/karolpiczak/ESC-50}, with human speech and speech-like sounds (e.g., babble, restaurant, bus, breathing, laughing,...) excluded. The reason for the exclusion was to ensure that the fine-tuned model does not learn to interpret the presence of multiple speakers as noise. 

The MS-SNSD recipe was edited so that each generated file contains noises from only one category (e.g., ``rain''), and for files generated from ASVSpoof data, the clean speech is only concatenated from samples of the same spoofing system.

For each of the three clean datasets listed above, we generated 20 hours of noisy speech (split 90:10 into training and validation sets). The total number of generated files in the training set was 38295 (4255 clean files + 8 noisy versions of each, at signal-to-noise ratio (SNR) levels of -20, -10, 0, 10, 20, 30, 40, 50), with durations ranging from 3 to 19 seconds. 

The generated clean and noisy data were only used for the first step of model fine-tuning. The training labels for this step were assigned automatically, based on a) SNR levels (for BAK prediction) or b) whether the original clean file was natural speech or synthesized (for SIG prediction). For BAK prediction, we used all the 38295 generated files, as described above. For SIG prediction, the setup was slightly different and will be explained in Section~\ref{sec:SIG}.

The second step of fine-tuning was done using only the VMC 2024 training and validation data: 
During the development phase of the challenge, while selecting the best model architecture and parameters, we only used the training set for fine-tuning; for the final submission on the evaluation set, we fine-tuned a new set of models using both the training and validation data, as permitted by the rules of the challenge. 

\subsection{Pretraining wav2vec 2.0 with degraded audio data}
\label{sec:w2v2-dgrd}

In our initial experiments during VMC 2024 itself, we used two existing pretrained  wav2vec 2.0 models, and found that the ``ClTRUS'' model offered slightly better performance at BAK prediction (evaluating the level of noise) than the basic wav2vec2-base. We attributed this to the richer and noisier training data used to pretrain the model.

After the challenge, we decided to apply this observation to SIG by creating a new pretrained wav2vec 2.0 model, using artificially degraded training data to simulate the possible signal distortions that can result from audio denoising and speech enhancement. This is a similar idea to the approach used by VMC 2024 team T03~\cite{coldenhoff2024}, who also generated artificially degraded data for pretraining, though for an otherwise very different system.

To generate the pretraining data, we used audio samples from the LibriSpeech (960 hours), AMI Meeting Corpus\footnote{https://groups.inf.ed.ac.uk/ami/corpus/} (cca 28 hours, using only the first 10 minutes of each ``Audio.Mix'' file), and CALLHOME American English\footnote{https://catalog.ldc.upenn.edu/LDC97S42} (cca 18 hours)  datasets, and artificially degraded them using a combination of one or more of the following techniques: 
\begin{itemize}
    \item no modification
    \item adding noise using the Lhotse toolkit~\cite{zelasko2021lhotse} (same noises as in Section \ref{sec:traindata}, plus noise samples from the MUSAN~\cite{snyder2015musanmusicspeechnoise} dataset and non-speech samples from the AudioSet~\cite{AudioSet2017} dataset\footnote{Downloaded from https://huggingface.co/datasets/confit/audioset-full (audio data) and https://research.google.com/audioset/download.html (labels). We used the "balanced\_train" set and excluded classes 0--37 (i.e., speech).}, at random SNR from -20 to 50)
    \item adding random reverberation using Lhotse
    \item applying speech enhancement or dereverberation\footnote{List of all models used (available on the HuggingFace Hub): speechbrain/sepformer-dns4-16k-enhancement, nvidia/se\_der\_sb\_16k\_small, nvidia/se\_den\_sb\_16k\_small, speechbrain/sepformer-whamr-enhancement, speechbrain/metricgan-plus-voicebank}: WPE dereverberation~\cite{drude2018nara} using Lhotse, Sepformer~\cite{subakan2021attention} speech enhancement, the Python package \texttt{audio-denoiser}, MetricGAN+ \cite{fu21_interspeech}, or Schrö\-dinger Bridge~\cite{jukic24_interspeech} denoising or dereverberation
    \item converting the data to a lossy audio format and back: MP3, GSM, Windows Media Audio (WMA), and Adaptive Multi-Rate (AMR-NB) audio codecs 
\end{itemize}

In total, the training set consisted of 299,972 utterances with a total duration of 1054 hours. Apart from the data, the wav2vec 2.0 model was pretrained using the same setup and settings as used both by the original wav2vec2-base~\cite{baevski2020wav2vec} and by ClTRUS~\cite{IS22_ASR_wav2vec2_Lehecka}.

\subsection{BAK prediction model (``SNR2BAK'')} 
\label{sec:BAK}

For the prediction of BAK scores, which measure the intrusiveness of background noise, we found the closest approximation in automatic \emph{signal-to-noise ratio (SNR) prediction}. Thus, in the first stage of fine-tuning, the model was fine-tuned to predict the level of added noise in the generated noisy data that was described in Section~\ref{sec:traindata}. However, to better approximate BAK scores, the original SNR values were first mapped to a range of 1--5 such that, for example, an SNR of -20 was converted to a label of 1.0 and an SNR of 50 to 4.5: 
\begingroup
\allowdisplaybreaks
\begin{align*}
    \text{noisy speech (SNR -20 to 50)} &\quad\rightarrow\quad y_i = 2 + 0.05 \cdot \text{SNR}_i\\ 
    \text{clean speech} &\quad\rightarrow\quad y_i=5.0
\end{align*}
\endgroup

In the second stage, we fine-tuned the model for BAK prediction on the labeled VMC 2024 data. 

\subsection{SIG prediction model (``spoof2SIG'')} 
\label{sec:SIG}

For SIG, which seems fairly equivalent to MOS prediction of synthetic speech, we decided to follow the example of another VMC 2022 competitor~\cite{stan22_interspeech} and start with the task of synthetic speech \emph{detection} -- fine-tuning a model to distinguish between natural and spoofed speech in the hope that the model would then carry some useful information over to SIG prediction. 

For this, we again used generated noisy data. In our original VMC 2024 submission, we only used data generated from the ASVSpoof 2019 and LibriSpeech datasets (excluding the default MS-SNSD data for a better balance between labels). ``Spoofed speech'' consisted of simulated data generated from the ASVSpoof 2019 ``spoofed'' (i.e., synthetic) utterances, while ``natural speech'' included everything else. For simplicity, the model was also fine-tuned for regression, with spoofed speech labeled as 1.0 and natural speech labeled as 5.0 to, again, better match the target range of SIG scores (we also tried fine-tuning for spoofing system classification, but it did not lead to better results on the VMC 2024 validation set). 

In the updated version of our models, we have modified this setup. Now, we use the full generated dataset from Section~\ref{sec:traindata}, and additionally also include some of the degraded data described in Section~\ref{sec:w2v2-dgrd}, which had been created for wav2vec 2.0 pretraining (specifically selecting utterances from the validation set). 
For the latter data, the labels are assigned as follows: 
clean, noisy or reverberated utterances are considered ``natural speech'', with a label of 5.0. 
Enhanced or dereverberated utterances are treated as ``spoofed'', with a label of 1.0. Utterances that have been degraded by converting them to a different codec are not included, as it is not clear what label they should be given in this context.

In the second stage, we fine-tuned the model for SIG prediction on the labeled VMC 2024 data. 
\subsection{OVRL prediction}
\label{sec:OVRL}

For OVRL prediction, we have considered two approaches. Initially, during VMC 2024 itself, the OVRL score was obtained as a simple average of the BAK and SIG scores (``OVRL$_A$'' in Figure~\ref{fig:voicemos2024_track3}).

Since then, we have added a dedicated OVRL prediction approach, which employs the fine-tuned BAK and SIG prediction models: we take the two fully fine-tuned models, without the prediction heads, freeze them completely, and add a new linear layer (consisting of one neuron) 
using outputs from both. This new layer alone is then trained for 200 epochs to predict the OVRL scores on the VMC24 training and/or validation data. In Figure~\ref{fig:voicemos2024_track3}, this approach is labeled as ``OVRL$_P$''.

\section{Results and conclusions}

In line with the official VMC 2024 protocol, system performance is reported using the linear (Pearson) correlation coefficient (LCC) between predictions and ground-truth scores. LCC is arguably the most useful measure in this context, as it captures how well predictions align with the subjective scale of human judgments and allows consistent comparison across systems.

\begin{figure}[ht]
    \centering
    \includegraphics[width=\linewidth]{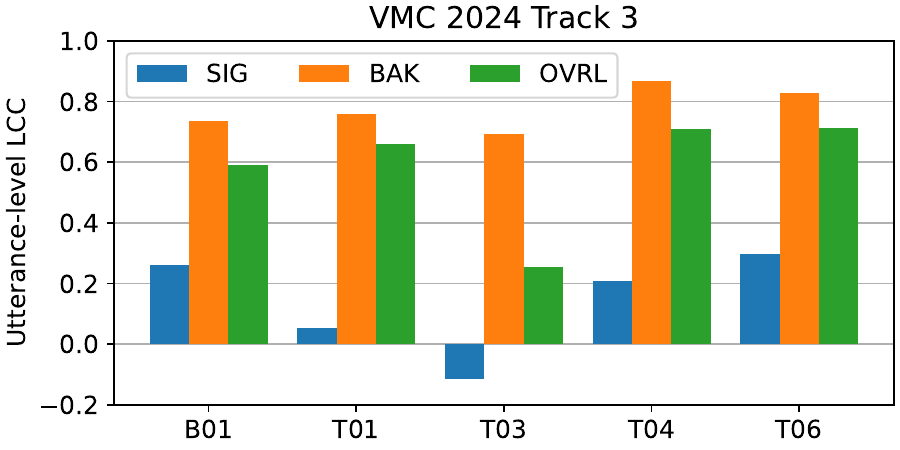}
    \caption{Official results of VMC 2024 Track 3. Our team is T04.}
    \label{fig:ofic_results_vmc24_track3}
    \vspace{-1em}
\end{figure}

The official results~\cite{huang2024voicemos2024} for all participating teams are shown in Figure~\ref{fig:ofic_results_vmc24_track3}. 
The results of our own submission (as Team T04, using a BAK prediction model fine-tuned from ClTRUS and a SIG prediction model fine-tuned from wav2vec2-base) are also
listed at the bottom of Table~\ref{tab:results_vm24_track3}, together with the other top-scoring team -- T06. 
In the rest of the table, we present the results of the extended system, with the different pretrained model combinations we have explored.

\begin{table*}[ht]
\centering

\caption{Results (as LCC) on Track 3 of VMC 2024 and on the CHiME 7 - UDASE dataset  (excluding files used in the VMC 2024 training and validation sets). ``f.-t. d.'': stage 2 fine-tuning data. }

  \label{tab:results_vm24_track3}
  \centering
  \small
  \begin{tabular}{r l *{12}{S[round-mode=places,table-format=0.3, round-precision=3,detect-weight,print-zero-integer=false]}}  
    \toprule

    & & \multicolumn{4}{c}{\small VMC 2024 validation set} & \multicolumn{4}{c}{\small VMC 2024 evaluation set} & \multicolumn{4}{c}{\small CHiME 7-UDASE (w/o VMC)}\\
    & & \multicolumn{4}{c}{\small (f.-t. d.: VMC train. set)} & \multicolumn{4}{c}{\small (f.-t. d.: VMC train. + valid. s.)}  & \multicolumn{4}{c}{\small (f.-t. d.: VMC train. + valid. s.)}\\

    \cmidrule(lr){3-6}
    \cmidrule(lr){7-10}
    \cmidrule(lr){11-14}

    \multicolumn{2}{c}{\small model combination} & & & \multicolumn{2}{c}{\footnotesize OVRL} & & & \multicolumn{2}{c}{\footnotesize OVRL} & & & \multicolumn{2}{c}{\footnotesize OVRL} \\
    \cmidrule(lr){1-2}
    \cmidrule(lr){5-6}
    \cmidrule(lr){9-10}
    \cmidrule(lr){13-14}

    {\footnotesize BAK} & {\footnotesize SIG} & {\footnotesize BAK} & {\footnotesize SIG} & {\footnotesize A} & {\footnotesize P} & {\footnotesize BAK} & {\footnotesize SIG} & {\footnotesize A} & {\footnotesize P} & {\footnotesize BAK} & {\footnotesize SIG} & {\footnotesize A} & {\footnotesize P}\\
                                         
    \midrule

    ClTRUS & ClTRUS  &  
        0.7753946967 & 0.8103854371 & 0.7922026724 & 0.8095384692 &  
        \B 0.8774719147 & 0.2964214092 & 0.7376090828 & 0.7276026704 & 
        0.8392234463 & \B 0.7664989889 & 0.7235093653 & 0.7499759917\\
    \B ClTRUS & \B w2v2-base  & 
         0.7753946967 & 0.8138131925 & \B 0.8311018638 & \B 0.8547094795 &
         \B 0.8774719147 & \B 0.5157985338 & \B 0.7806803535 & \B 0.7759216939 &
         0.8392234463 & 0.7257709094 & 0.7140265858 &  0.729919719\\
    ClTRUS & w2v2-dgrd & 
         0.7753946967 & \B 0.8401500198 & 0.7761928306 & 0.80075996 &
         \B 0.8774719147 & 0.4789033966 & 0.7463224781 & 0.7365751199 &
         0.8392234463 & 0.6971587869 & 0.6729496395 & 0.691732967 \\
    w2v2-base & ClTRUS  & 
        0.7888961749 & 0.8103854371 & 0.7458369768 & 0.7595277266 &
        0.8452824897 & 0.2964214092 & 0.7078447911 & 0.6948858526 &
        0.8042945972 & \B 0.7664989889 & 0.7019340282 & 0.7275652148\\
    w2v2-base & w2v2-base & 
        0.7888961749 & 0.8138131925 & 0.7833132609 & 0.8095333319 &
        0.8452824897 & \B 0.5157985338 & 0.7519111108 & 0.7375325669 &
        0.8042945972 & 0.7257709094 & 0.6883520748 & 0.7014901145 \\
    w2v2-base & w2v2-dgrd  & 
        0.7888961749 & \B 0.8401500198 & 0.7267850233 & 0.7440406662 &
        0.8452824897 & 0.4789033966 & 0.7148779106 & 0.6944453739 &
        0.8042945972 & 0.6971587869 & 0.6578636237 & 0.6800786071\\
    w2v2-dgrd & ClTRUS  & 
         \B 0.8030678001 & 0.8103854371 & 0.7488149575 & 0.7749766473 &
         0.8676125477 & 0.2964214092 & 0.711208316 & 0.6951351687 &
         \B 0.8602148946 & \B 0.7664989889 & \B 0.7461767864 & \B 0.769387781\\
    w2v2-dgrd & w2v2-base  & 
         \B 0.8030678001 & 0.8138131925 & 0.7832729763 & 0.823031455 &
         0.8676125477 & \B 0.5157985338 & 0.7632294458 & 0.7499834388 &
         \B 0.8602148946 & 0.7257709094 & 0.7343461955 & 0.7444340399\\
    w2v2-dgrd & w2v2-dgrd  & 
         \B 0.8030678001 & \B 0.8401500198 & 0.7317718442 & 0.7617473421 &
         0.8676125477 & 0.4789033966 & 0.7195255654 & 0.6914952498  &
         \B 0.8602148946 & 0.6971587869 & 0.701365811 & 0.7134241298 \\

    \midrule

    \multicolumn{2}{l}{Our VMC 2024 submission (T04)}  & 
        0.7975776981 & 0.7539179817 & 0.7220835347 & {--}  &
         0.8671628269 & 0.206910607 & 0.7110318864 & {--}  &
        0.8188463072 & 0.68410354 & 0.5948944936 & {--}\\ 

    \multicolumn{2}{l}{Team T06 (1st/2nd place)} & {--} & {--} & {--} & {--} & 0.827231 & 0.297102 & \multicolumn{2}{S[round-mode=places,table-format=0.3, round-precision=3,detect-weight,print-zero-integer=false]}{0.712608} & {--} & {--} & {--} & {--}\\

    \bottomrule
  \end{tabular}

  \vspace{-1em}
  
\end{table*}

Note that in the course of updating our results, we retrained all models from scratch, with slightly different settings than previously (mainly a larger batch size and number of epochs). This has caused the final values to change compared to our original submission, even for the BAK models, where training data remained the same.

As previously mentioned, the VMC 2024 evaluation data did not share the same source as the training and validation data -- the latter two originating from the CHiME 7 -- UDASE \cite{LEGLAIVE2025101685} dataset. For comparison, we have decided to also test our final models on this same dataset, using the remaining audio samples that had not been used\footnote{The shared files were identified by manual comparison and are listed in the appendix of the arXiv version of this paper (i.e., in this document)} in VMC 2024. 
The results on the remaining 588 CHiME~7 -- UDASE files are also shown in Table~\ref{tab:results_vm24_track3}. 

\subsection{Discussion and conclusions}

In interpreting our results, it is important to note that the training data for VMC 2024 Track 3 was extremely limited, which naturally introduces some variability in evaluation. Across repeated fine-tunings with slightly different settings (e.g., number of epochs or checkpoint selection), we observed fluctuations of up to $\pm0.050$ in LCC. This variation is modest and does not change the overall conclusions, but it does mean that differences between the various model combinations -- which often fall within this range -- should be interpreted cautiously. 

With this in mind, some clear tendencies can still be identified. The newly pretrained w2v2-dgrd model did not consistently outperform the other wav2vec 2.0 variants. However, the artificially degraded data used in its pretraining turned out to be highly useful when included in the first fine-tuning stage for the SIG predictor, leading to a marked improvement from 0.207 in our original submission to 0.516. While SIG remains more challenging to predict than BAK, this represents a substantial gain and clearly surpasses the results of all original VMC 2024 submissions.

The BAK predictor continues to perform very strongly, showing high correlations with the ground truth across all tested conditions and again outperforming all other participating teams. This confirms that approximating BAK through SNR prediction is an effective and reliable strategy in this task.

For OVRL, the simple averaging of the predicted SIG and BAK scores proved very effective -- and also unexpectedly appropriate, as post-challenge analysis reveals a near-perfect correlation ($\text{LCC}=0.985$) between the ground-truth OVRL scores and the averages of the ground-truth SIG and BAK scores, confirming the suitability of our choice. A dedicated OVRL model gave similar performance, with slightly higher correlations on in-domain data (VMC validation and CHiME 7–UDASE), but slightly lower values on the evaluation set.

Overall, our approach, though deliberately simple in design, has demonstrated that transfer learning combined with carefully generated auxiliary data can achieve strong and reliable predictions of P.835 metrics under extremely limited data conditions. We expect that further improvements will depend less on architectural changes and more on access to larger subjectively labeled datasets, which would allow clearer differentiation between pretrained models and more detailed error analyses.

\subsubsection*{Declarations}

\textbf{Funding Acknowledgments:} 
This work was created with the support of the project ``R\&D of Technologies for Advanced Digitization in the Pilsen Metropolitan Area (DigiTech)'' No.: CZ.02.01.01/00/\allowbreak{}23\_021/0008436 co-financed by the European Union.    
Computational resources were provided by the e-INFRA CZ project (ID:90254), supported by the Ministry of Education, Youth and Sports of the Czech Republic.
The authors have no other relevant financial or nonfinancial interests to disclose.

\noindent
\textbf{Compliance with Ethical Standards:} 
This study did not involve any human or animal subjects, so no ethical approval was required.

\FloatBarrier
\bibliographystyle{IEEEtran}
\bibliography{mybib}

\begin{thebibliography}{10}
\providecommand{\url}[1]{#1}
\csname url@samestyle\endcsname
\providecommand{\newblock}{\relax}
\providecommand{\bibinfo}[2]{#2}
\providecommand{\BIBentrySTDinterwordspacing}{\spaceskip=0pt\relax}
\providecommand{\BIBentryALTinterwordstretchfactor}{4}
\providecommand{\BIBentryALTinterwordspacing}{\spaceskip=\fontdimen2\font plus
\BIBentryALTinterwordstretchfactor\fontdimen3\font minus \fontdimen4\font\relax}
\providecommand{\BIBforeignlanguage}[2]{{%
\expandafter\ifx\csname l@#1\endcsname\relax
\typeout{** WARNING: IEEEtran.bst: No hyphenation pattern has been}%
\typeout{** loaded for the language `#1'. Using the pattern for}%
\typeout{** the default language instead.}%
\else
\language=\csname l@#1\endcsname
\fi
#2}}
\providecommand{\BIBdecl}{\relax}
\BIBdecl

\bibitem{Wester2015}
M.~Wester, C.~Valentini-Botinhao, and G.~E. Henter, ``{Are we using enough listeners? No! -- An empirically-supported critique of Interspeech 2014 TTS evaluations},'' in \emph{Proc. Interspeech}, 2015, pp. 3476--3480.

\bibitem{huang2024voicemos2024}
W.-C. Huang \emph{et~al.}, ``The {VoiceMOS Challenge} 2024: Beyond speech quality prediction,'' in \emph{Proc. SLT}, 2024, pp. 803--810.

\bibitem{Kunesova2025VoiceMOS_journal}
M.~Kune{\v s}ov{\'a}, J.~Matou{\v s}ek, J.~Lehe{\v c}ka, J.~{\v S}vec, D.~Tihelka, and Z.~Hanzl{\'i\v c}ek, ``Three years of {VoiceMOS} challenges: Lessons learned by the {UWB-NTIS-TTS} team,'' \emph{IEEE Access}, vol.~13, pp. 140\,152--140\,174, 2025.

\bibitem{ITU.P.800.1}
{ITU-T Recommendation P.800.1}, ``Mean opinion score {(MOS)} terminology,'' Int. Telecom. Union, Tech. Rep., 2003.

\bibitem{ITU.P.835}
{ITU-T Recommendation P.835}, ``Subjective test methodology for evaluating speech communication systems that include noise suppression algorithm,'' Int. Telecom. Union, Tech. Rep., 2003.

\bibitem{LEGLAIVE2025101685}
S.~Leglaive \emph{et~al.}, ``Objective and subjective evaluation of speech enhancement methods in the {UDASE} task of the 7th {CHiME} challenge,'' \emph{Comp. Speech \& Lang.}, vol.~89, p. 101685, 2025.

\bibitem{valentinibotinhao16_ssw}
C.~Valentini-Botinhao, X.~Wang, S.~Takaki, and J.~Yamagishi, ``Investigating {RNN-based} speech enhancement methods for noise-robust text-to-speech,'' in \emph{9th ISCA Workshop on Speech Synthesis Workshop (SSW 9)}, 2016, pp. 146--152.

\bibitem{Rix2001}
A.~W. Rix, J.~G. Beerends, M.~P. Hollier, and A.~P. Hekstra, ``{Perceptual evaluation of speech quality (PESQ) - A new method for speech quality assessment of telephone networks and codecs},'' in \emph{Proc. ICASSP}, 2001, pp. 749--752.

\bibitem{ITU.P.863}
{ITU-T Recommendation P.863}, ``Perceptual objective listening quality assessment,'' Int. Telecom. Union, Tech. Rep., 2018.

\bibitem{Patton2016}
B.~Patton, Y.~Agiomyrgiannakis, M.~Terry, K.~Wilson, R.~A. Saurous, and D.~Sculley, ``{AutoMOS: Learning a non-intrusive assessor of naturalness-of-speech},'' in \emph{Int. Conf. Neural Inf. Process. Syst.}, 2016.

\bibitem{Yoshimura2016}
T.~Yoshimura, G.~E. Henter, O.~Watts, M.~Wester, J.~Yamagishi, and K.~Tokuda, ``{A hierarchical predictor of synthetic speech naturalness using neural networks},'' in \emph{Proc. Interspeech}, 2016, pp. 342--346.

\bibitem{Saeki22c_UTMOS}
T.~Saeki, D.~Xin, W.~Nakata, T.~Koriyama, S.~Takamichi, and H.~Saruwatari, ``{UTMOS}: {UTokyo-SaruLab} system for {VoiceMOS Challenge} 2022,'' in \emph{Interspeech}, 2022, pp. 4521--4525.

\bibitem{Avila2019}
A.~R. Avila, H.~Gamper, C.~Reddy, R.~Cutler, I.~Tashev, and J.~Gehrke, ``Non-intrusive speech quality assessment using neural networks,'' in \emph{Proc. ICASSP}, 2019, pp. 631--635.

\bibitem{Dong2020}
X.~Dong and D.~S. Williamson, ``An attention enhanced multi-task model for objective speech assessment in real-world environments,'' in \emph{Proc. ICASSP}, 2020, pp. 911--915.

\bibitem{deoliveira25_interspeech}
D.~{de Oliveira}, J.~Richter, J.-M. Lemercier, S.~Welker, and T.~Gerkmann, ``Non-intrusive speech quality assessment with diffusion models trained on clean speech,'' in \emph{Proc. Interspeech}, 2025, pp. 2330--2334.

\bibitem{mittag21_interspeech}
G.~Mittag, B.~Naderi, A.~Chehadi, and S.~Möller, ``{NISQA}: A deep {CNN}-self-attention model for multidimensional speech quality prediction with crowdsourced datasets,'' in \emph{Proc. Interspeech}, 2021, pp. 2127--2131.

\bibitem{wardah25_interspeech}
W.~Wardah \emph{et~al.}, ``{SQ-AST:} a transformer-based model for speech quality prediction,'' in \emph{Proc. Interspeech}, 2025, pp. 2335--2339.

\bibitem{Reddy2022_DNSMOS}
C.~K.~A. Reddy, V.~Gopal, and R.~Cutler, ``{DNSMOS P.835}: A non-intrusive perceptual objective speech quality metric to evaluate noise suppressors,'' in \emph{ICASSP}, 2022, pp. 886--890.

\bibitem{baevski2020wav2vec}
A.~Baevski, Y.~Zhou, A.~Mohamed, and M.~Auli, ``wav2vec 2.0: A framework for self-supervised learning of speech representations,'' \emph{Adv. Neural Inf. Process. Syst.}, vol.~33, pp. 12\,449--12\,460, 2020.

\bibitem{IS22_ASR_wav2vec2_Lehecka}
J.~Lehe{\v{c}}ka, J.~{\v{S}}vec, A.~Pra{\v{z}}{\'{a}}k, and J.~V. Psutka, ``Exploring capabilities of monolingual audio transformers using large datasets in automatic speech recognition of {Czech},'' in \emph{Proc. Interspeech}, 2022, pp. 1831--1835.

\bibitem{reddy2019scalable}
C.~K. Reddy, E.~Beyrami, J.~Pool, R.~Cutler, S.~Srinivasan, and J.~Gehrke, ``A scalable noisy speech dataset and online subjective test framework,'' \emph{Proc. Interspeech}, pp. 1816--1820, 2019.

\bibitem{pirker11_ptdbtug}
G.~Pirker, M.~Wohlmayr, S.~Petrik, and F.~Pernkopf, ``A pitch tracking corpus with evaluation on multipitch tracking scenario,'' in \emph{Proc. Interspeech}, 2011, pp. 1509--1512.

\bibitem{Panayotov_Librispeech}
V.~Panayotov, G.~Chen, D.~Povey, and S.~Khudanpur, ``{LibriSpeech:} an {ASR} corpus based on public domain audio books,'' in \emph{Proc. ICASSP}, 2015, pp. 5206--5210.

\bibitem{Wang_ASVspoof2019}
X.~Wang \emph{et~al.}, ``{ASVspoof 2019:} a large-scale public database of synthesized, converted and replayed speech,'' \emph{Comp. Speech \& Lang.}, vol.~64, p. 101114, 2020.

\bibitem{coldenhoff2024}
J.~Coldenhoff, ``Objective perception metrics for audio quality,'' Master's thesis, École Polytechnique Fédérale de Lausanne, 2024.

\bibitem{zelasko2021lhotse}
\BIBentryALTinterwordspacing
P.~Żelasko, D.~Povey, J.~Trmal, and S.~Khudanpur, ``Lhotse: a speech data representation library for the modern deep learning ecosystem,'' 2021. [Online]. Available: \url{https://arxiv.org/abs/2110.12561}
\BIBentrySTDinterwordspacing

\bibitem{snyder2015musanmusicspeechnoise}
\BIBentryALTinterwordspacing
D.~Snyder, G.~Chen, and D.~Povey, ``{MUSAN}: A music, speech, and noise corpus,'' 2015. [Online]. Available: \url{https://arxiv.org/abs/1510.08484}
\BIBentrySTDinterwordspacing

\bibitem{AudioSet2017}
J.~F. Gemmeke \emph{et~al.}, ``{Audio Set}: An ontology and human-labeled dataset for audio events,'' in \emph{Proc. ICASSP}, 2017, pp. 776--780.

\bibitem{drude2018nara}
L.~Drude, J.~Heymann, C.~Boeddeker, and R.~Haeb-Umbach, ``{NARA-WPE}: A {Python} package for weighted prediction error dereverberation in {Numpy} and {Tensorflow} for online and offline processing,'' in \emph{Sp. Comm.; 13th ITG-Symp.}, 2018, pp. 1--5.

\bibitem{subakan2021attention}
C.~Subakan, M.~Ravanelli, S.~Cornell, M.~Bronzi, and J.~Zhong, ``Attention is all you need in speech separation,'' in \emph{Proc. ICASSP}, 2021.

\bibitem{fu21_interspeech}
S.-W. Fu \emph{et~al.}, ``{MetricGAN+}: An improved version of {MetricGAN} for speech enhancement,'' in \emph{Proc. Interspeech}, 2021, pp. 201--205.

\bibitem{jukic24_interspeech}
A.~Jukić, R.~Korostik, J.~Balam, and B.~Ginsburg, ``Schrödinger bridge for generative speech enhancement,'' in \emph{Proc. Interspeech}, 2024, pp. 1175--1179.

\bibitem{stan22_interspeech}
A.~Stan, ``The {ZevoMOS} entry to {VoiceMOS Challenge} 2022,'' in \emph{Proc. Interspeech}, 2022, pp. 4516--4520.

\end{thebibliography}

\newpage
\appendix

\section{List of excluded CHiME7 - UDASE files}

The following files from the CHiME7 - UDASE dataset were excluded from the evaluation in Table~\ref{tab:results_vm24_track3}, as they are present in the VoiceMOS 2024 Track 3 training and validation sets. (These lists were compiled through a manual comparison of the three datasets.)

\vspace{10pt}
\textbf{Used in the VMC 2024 training set (60 files):}\\
C0/S01\_P02\_8\_output.wav, C0/S01\_P03\_12\_output.wav, C0/S01\_P04\_25\_output.wav, C0/S01\_P04\_39\_output.wav, C0/S01\_P04\_41\_output.wav, C0/S21\_P45\_15\_output.wav, C0/S21\_P45\_22\_output.wav, C0/S21\_P45\_5\_output.wav, C0/S21\_P46\_10\_output.wav, C0/S21\_P47\_20\_output.wav, C0/S21\_P47\_5\_output.wav, C0/S21\_P47\_7\_output.wav, C1/S01\_P02\_8\_output.wav, C1/S01\_P03\_12\_output.wav, C1/S01\_P04\_25\_output.wav, C1/S01\_P04\_39\_output.wav, C1/S01\_P04\_41\_output.wav, C1/S21\_P45\_15\_output.wav, C1/S21\_P45\_22\_output.wav, C1/S21\_P45\_5\_output.wav, C1/S21\_P46\_10\_output.wav, C1/S21\_P47\_20\_output.wav, C1/S21\_P47\_5\_output.wav, C1/S21\_P47\_7\_output.wav, C2/S01\_P02\_8\_output.wav, C2/S01\_P03\_12\_output.wav, C2/S01\_P04\_25\_output.wav, C2/S01\_P04\_39\_output.wav, C2/S01\_P04\_41\_output.wav, C2/S21\_P45\_15\_output.wav, C2/S21\_P45\_22\_output.wav, C2/S21\_P45\_5\_output.wav, C2/S21\_P46\_10\_output.wav, C2/S21\_P47\_20\_output.wav, C2/S21\_P47\_5\_output.wav, C2/S21\_P47\_7\_output.wav, C3/S01\_P02\_8\_output.wav, C3/S01\_P03\_12\_output.wav, C3/S01\_P04\_25\_output.wav, C3/S01\_P04\_39\_output.wav, C3/S01\_P04\_41\_output.wav, C3/S21\_P45\_15\_output.wav, C3/S21\_P45\_22\_output.wav, C3/S21\_P45\_5\_output.wav, C3/S21\_P46\_10\_output.wav, C3/S21\_P47\_20\_output.wav, C3/S21\_P47\_5\_output.wav, C3/S21\_P47\_7\_output.wav, C4/S01\_P02\_8\_output.wav, C4/S01\_P03\_12\_output.wav, C4/S01\_P04\_25\_output.wav, C4/S01\_P04\_39\_output.wav, C4/S01\_P04\_41\_output.wav, C4/S21\_P45\_15\_output.wav, C4/S21\_P45\_22\_output.wav, C4/S21\_P45\_5\_output.wav, C4/S21\_P46\_10\_output.wav, C4/S21\_P47\_20\_output.wav, C4/S21\_P47\_5\_output.wav, C4/S21\_P47\_7\_output.wav

\vspace{10pt}
\textbf{Used in the VMC 2024 validation set (40 files):}\\
C0/S01\_P01\_18\_output.wav, C0/S01\_P02\_4\_output.wav, C0/S01\_P04\_0\_output.wav, C0/S01\_P04\_15\_output.wav, C0/S21\_P45\_45\_output.wav, C0/S21\_P46\_13\_output.wav, C0/S21\_P47\_27\_output.wav, C0/S21\_P47\_4\_output.wav, C1/S01\_P01\_18\_output.wav, C1/S01\_P02\_4\_output.wav, C1/S01\_P04\_0\_output.wav, C1/S01\_P04\_15\_output.wav, C1/S21\_P45\_45\_output.wav, C1/S21\_P46\_13\_output.wav, C1/S21\_P47\_27\_output.wav, C1/S21\_P47\_4\_output.wav, C2/S01\_P01\_18\_output.wav, C2/S01\_P02\_4\_output.wav, C2/S01\_P04\_0\_output.wav, C2/S01\_P04\_15\_output.wav, C2/S21\_P45\_45\_output.wav, C2/S21\_P46\_13\_output.wav, C2/S21\_P47\_27\_output.wav, C2/S21\_P47\_4\_output.wav, C3/S01\_P01\_18\_output.wav, C3/S01\_P02\_4\_output.wav, C3/S01\_P04\_0\_output.wav, C3/S01\_P04\_15\_output.wav, C3/S21\_P45\_45\_output.wav, C3/S21\_P46\_13\_output.wav, C3/S21\_P47\_27\_output.wav, C3/S21\_P47\_4\_output.wav, C4/S01\_P01\_18\_output.wav, C4/S01\_P02\_4\_output.wav, C4/S01\_P04\_0\_output.wav, C4/S01\_P04\_15\_output.wav, C4/S21\_P45\_45\_output.wav, C4/S21\_P46\_13\_output.wav, C4/S21\_P47\_27\_output.wav, C4/S21\_P47\_4\_output.wav

\vspace{10pt}

All other CHiME7 - UDASE files not listed here (588 audio files, including clean references) were used for the evaluation.

\end{document}